\documentclass[epj]{webofc}
\usepackage[varg]{txfonts}   
\usepackage{color}
\woctitle{6\textsuperscript{th} International conference on
Physics Opportunities at Electron-Ion colliders (POETIC6)}
\newcommand{\ncteq}{{\tt nCTEQ}}

\newcommand{\ncteqfit}{{\tt nCTEQ15}}
\newcommand{\GeV}{{\rm GeV}}

\begin{document}
\title{Frontiers of QCD with Precision nPDFs
\thanks{Based on work in collaboration with
D.~B.~Clark, 
E.~Godat,
T.~Ježo, 
C.~Keppel, 
K.~Kova\v{r}ík, 
J.G.~Morf{\'{i}}n, 
P.~Nadolsky,
J.F.~Owens, 
\&
J.Y.~Yu.
This work is supported in part by the U.S.~Department of Energy under grant
DE-FG02-04ER41299, 
and by Projet international de cooperation scientifique PICS05854 between
France and the U.S.
}
}
%
%

\author{
Aleksander Kusina\inst{1}\fnsep\thanks{\email{kusina@lpsc.in2p3.fr}}
\and
Florian Lyonnet\inst{2}\fnsep\thanks{\email{flyonnet@smu.edu}}
\and
Fredrick I. Olness\inst{2}\fnsep\thanks{\email{olness@smu.edu}}\fnsep\thanks{Presenter}
\and
Ingo Schienbein\inst{1}\fnsep\thanks{\email{schien@lpsc.in2p3.fr}}
}
\institute{
LPSC, Universit\'e Joseph Fourier/CNRS-IN2P3/INPG, UMR5821, Grenoble, F-38026, France
\and
Southern Methodist University, Dallas, TX 75275, USA}

\abstract{%
Searches for new physics will increasingly depend on identifying
deviations from precision Standard Model (SM) predictions.
Quantum Chromodynamics (QCD) will necessarily play a central role in
this endeavor as it provides the framework for the parton model.  
However, as we move to higher orders and into
extreme kinematic regions, we begin to see the full complexities of
the QCD theory.
Recent theoretical developments improve our ability to analyze both
proton and nuclear PDFs across the full kinematic range.
These developments are incorporated into the new \ncteqfit\ PDFs, and we
review these developments with respect to future measurements, and
identify areas where additional effort is required.
}
\maketitle

\section{The Character and Consequences of QCD\protect\footnote{In this 
short presentation we provide only an abbreviated set of references.}}
\label{intro}

``QCD  is our most perfect physical theory.'' \footnote{
{\it What QCD Tells Us About Nature -- and Why We Should Listen},
by  Frank Wilczek.\cite{Wilczek:1999id}
}
Its scope is wide ranging spanning  nuclear physics, particle physics,  cosmology, 
and astrophysics. It also contains a wealth of phenomena including: 
``radiative corrections, running couplings, confinement, 
spontaneous (chiral) symmetry breaking, anomalies, and instantons''\cite{Wilczek:1999id}.

QCD is both a force whose properties we want to study, and also a tool 
we can use to link experimentally measured hadronic cross sections to 
theoretically calculable cross sections in  the context of the parton model.
The basic statement  of the parton model is: 
$
\sigma_{A\to X} = f_{a/A} \otimes \widehat{\sigma}_{a\to X}
$
where we factorize the complex non-perturbative 
QCD interactions into the  Parton Distribution Functions (PDFs) $f_{a/A}$ and the 
theoretical partonic cross section $\widehat{\sigma}_{a\to X}$.
Here, 
$\sigma_{A\to X}$ is  the physically measurable cross section, and 
$A$ can represent either a proton or a nucleon. 

\begin{figure}[ht]
\centering
\includegraphics[width=0.75\textwidth,clip]{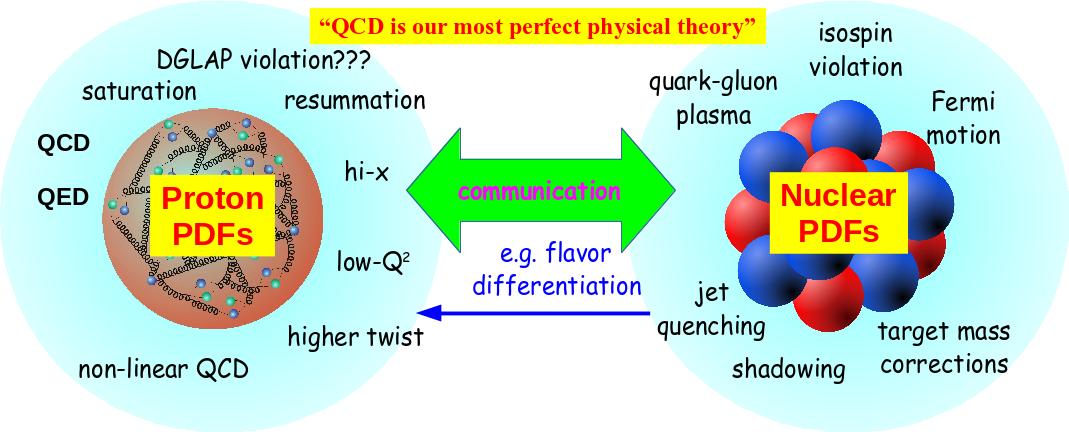}
\caption{We schematically represent the variety of phenomena which may enter
when considering proton and nuclear PDFs at high precision. 
We emphasize that the nuclear PDFs and corrections are important 
for flavor differentiation. 
}
\label{fig:sch}       
\end{figure}

As we push our calculations to higher precision and into extreme kinematic regions 
(such as low-$Q^2$ or hi/low-$x$) we may encounter a variety of phenomena such as 
those  illustrated schematically in Fig.\ref{fig:sch}. 
While many of these phenomena can be neglected at lower orders and lower  precision, 
if we are to properly analyze the new generation of high-precision high-statistics experiments, 
we must consider the potential impact of these manifestations of the QCD theory.

The recent performance of the LHC has exceeded expectations and
produced an unprecedented number of events to be analyzed. 
However, to advance our understanding of these processes and the inherent aspects of the QCD theory
 requires both 
improved theoretical tools {\it and} complementary data  from other experiments such as 
 an   Electron Ion Collider (EIC) and   Large Hadron  electron Collider (LHeC). 
In  the same manner that the combination of the 
HERA $ep$ and Tevatron $p\bar{p}$ data provided maximum knowledge of the underlying physics, 
the combination of the LHC and future lepton-hadron data can further expand our 
search for new physics and our understanding of QCD. 

To carry out this program requires both the above data, and also 
improved theoretical tools to perform the analysis. We briefly review 
some of the recent progress, and identify areas where additional effort is required.

\subsection{Nuclear PDFs}

Historically, the nuclear PDFs (nPDFs) and the associated nuclear correction factors
have played a key role in the flavor
differentiation of the proton PDFs.  The  common processes used to
extract the proton PDFs were Deeply Inelastic Scattering (DIS),
Drell-Yan (DY) lepton-hadron scattering, and jet production. These
 processes involve different linear combinations of the
partonic flavors, and the neutrino-nucleon DIS processes are
especially valuable as we can extract the four quantities $\{ F_2^{\nu
  N}$, $F_2^{\bar\nu N}$, $F_3^{\nu N}$, $F_3^{\bar\nu N} \}$. Due to
the small neutrino cross section, these measurements are usually
performed on heavy nuclear targets; hence, a detailed understanding
of the nPDFs and their corrections are essential to
constraining the flavor structure of both the nuclear and proton PDFs.

\begin{figure}[ht]
\centering
\includegraphics[width=0.50\textwidth,clip]{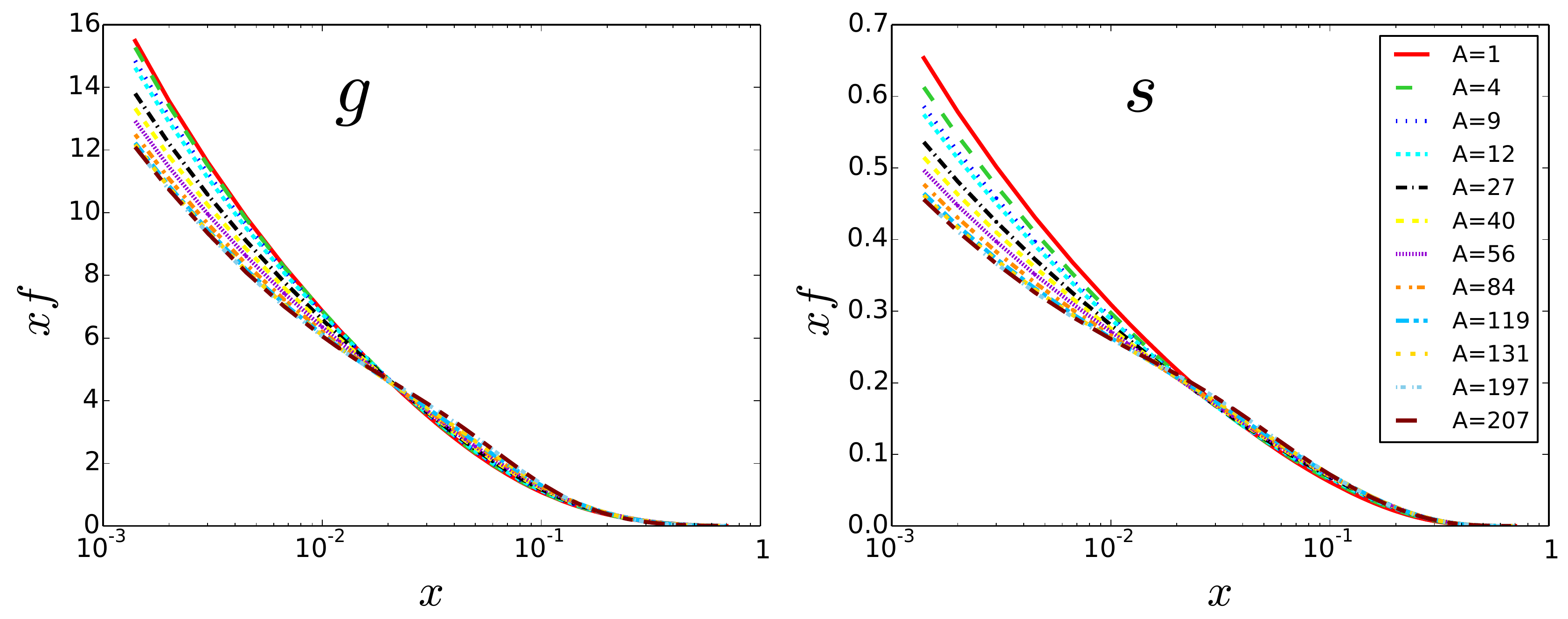}
\hfil
\includegraphics[width=0.45\textwidth,clip]{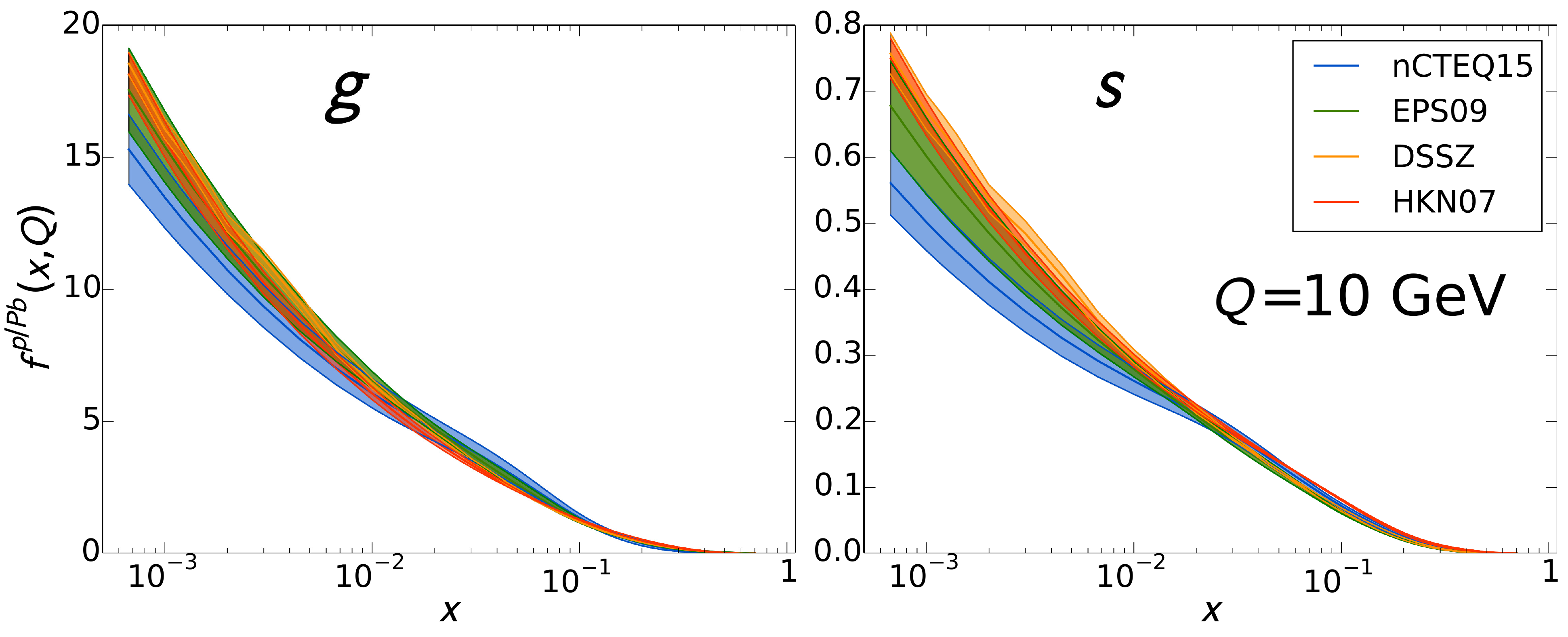}
\caption{
a) The nCTEQ15 PDFs  for a range of nuclear $A$ values.\cite{Kovarik:2015cma}
b) The nPDF  uncertainty bands, together with  HKN07,\cite{Hirai:2007sx}
EPS09,\cite{Eskola:2009uj} and DSSZ.\cite{deFlorian:2011fp}
}
\label{fig:ncteq}       
\end{figure}

To comprehensively analyze both proton and nuclear data in a single
unified framework, we have developed the \ncteq\ project which can
combine both data sets (including
uncertainties) into the fit.  The \ncteqfit\ set of 
nPDFs\cite{Kovarik:2015cma} is illustrated in
Fig.\ref{fig:ncteq}, and this extends the CTEQ proton PDFs fits to
include the nuclear dependence across the full $A$ range from proton up to
${}^{208}$Pb. 

\subsection{The  Strange Quark}

\begin{figure}[ht]
\centering
\includegraphics[width=0.40\textwidth,clip]{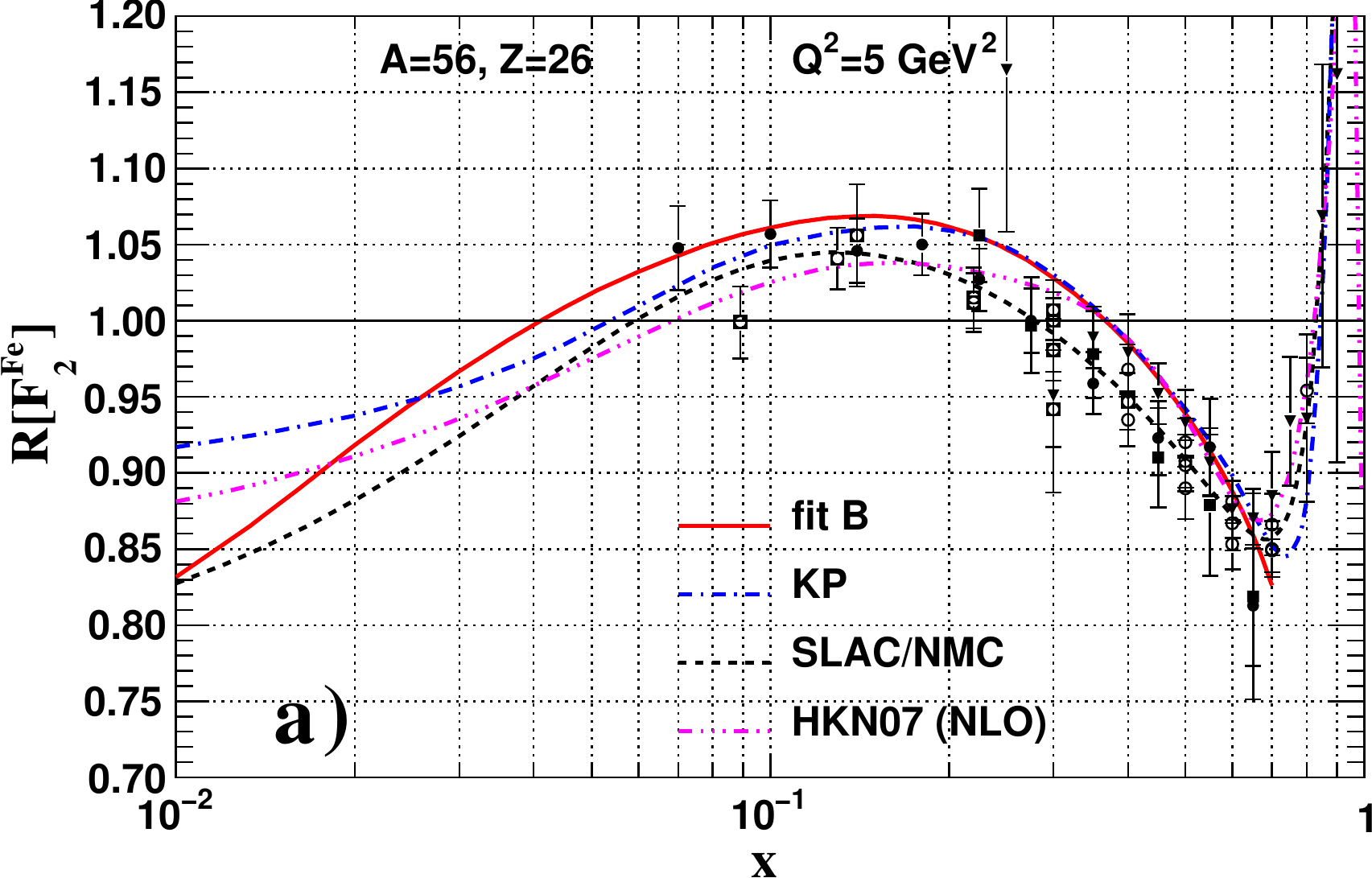}
\hfil
\includegraphics[width=0.40\textwidth,clip]{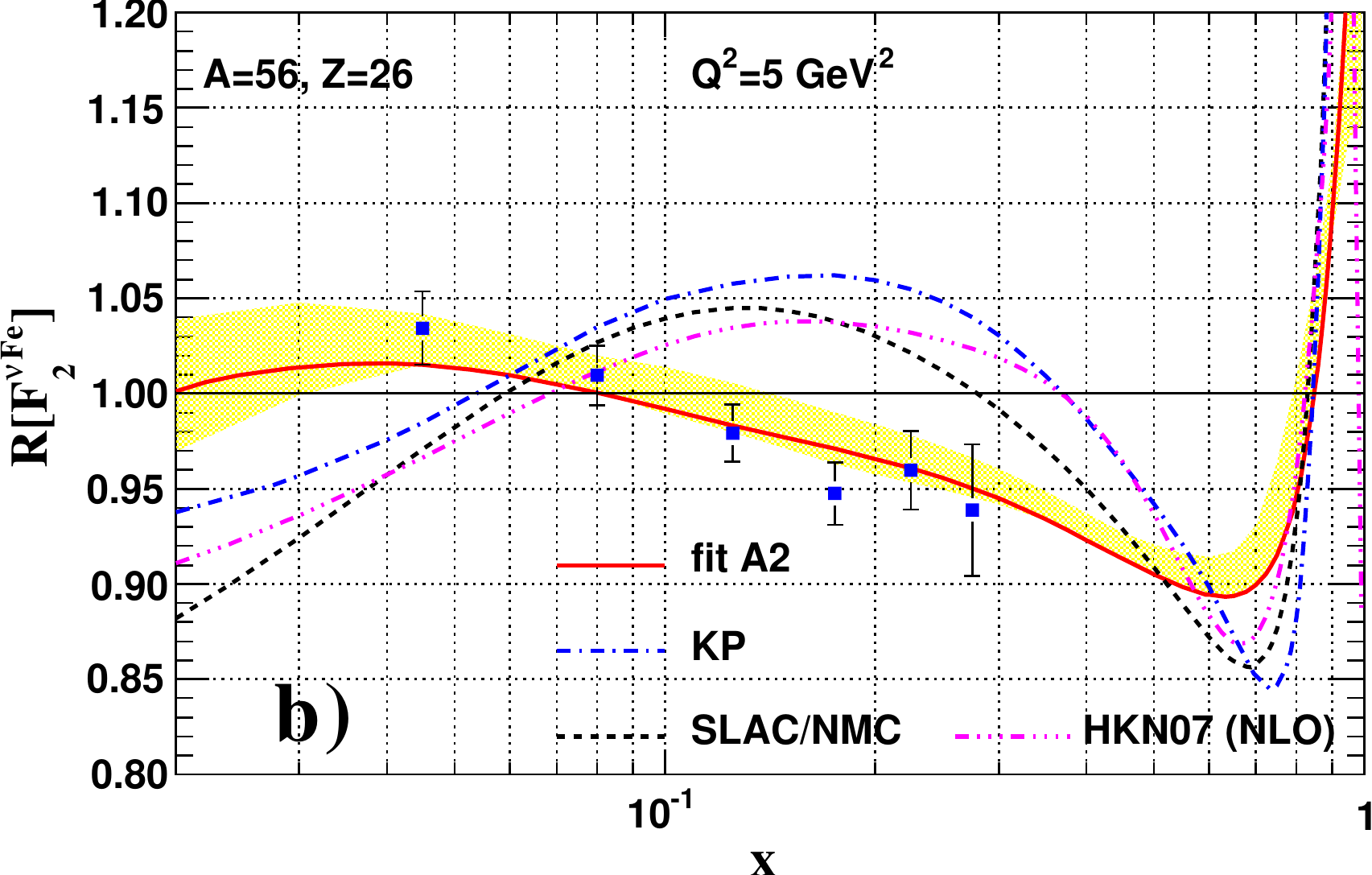}
\caption{
The computed nuclear correction ratio, $R=F_{2}^{Fe}/F_{2}^{D}$
as a function of $x$. 
Figure-a) shows the fit using the  neutral-current lepton ($\ell^{\pm}N$) DIS data (KP, SLAC/NMC,
HKN07). 
Figure-b) shows the fit using the  charged-current neutrino ($\nu N$) DIS data;
the data are from the NuTeV experiment. 
See Ref.~\cite{Kovarik:2010uv}
}
\label{fig:rf}       
\end{figure}

\begin{figure}[ht]
\centering
\includegraphics[width=0.55\textwidth,clip]{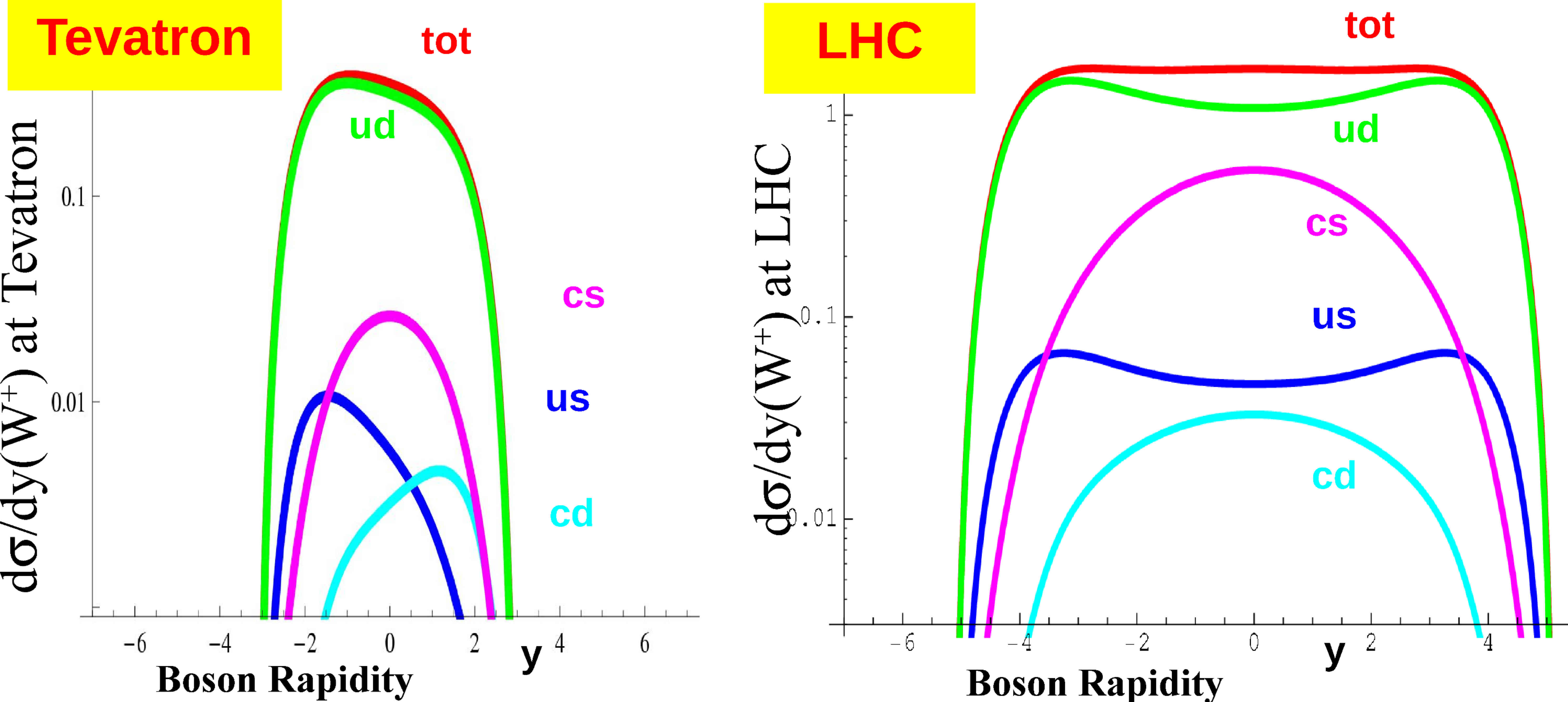}
\hfil
\includegraphics[width=0.30\textwidth,clip]{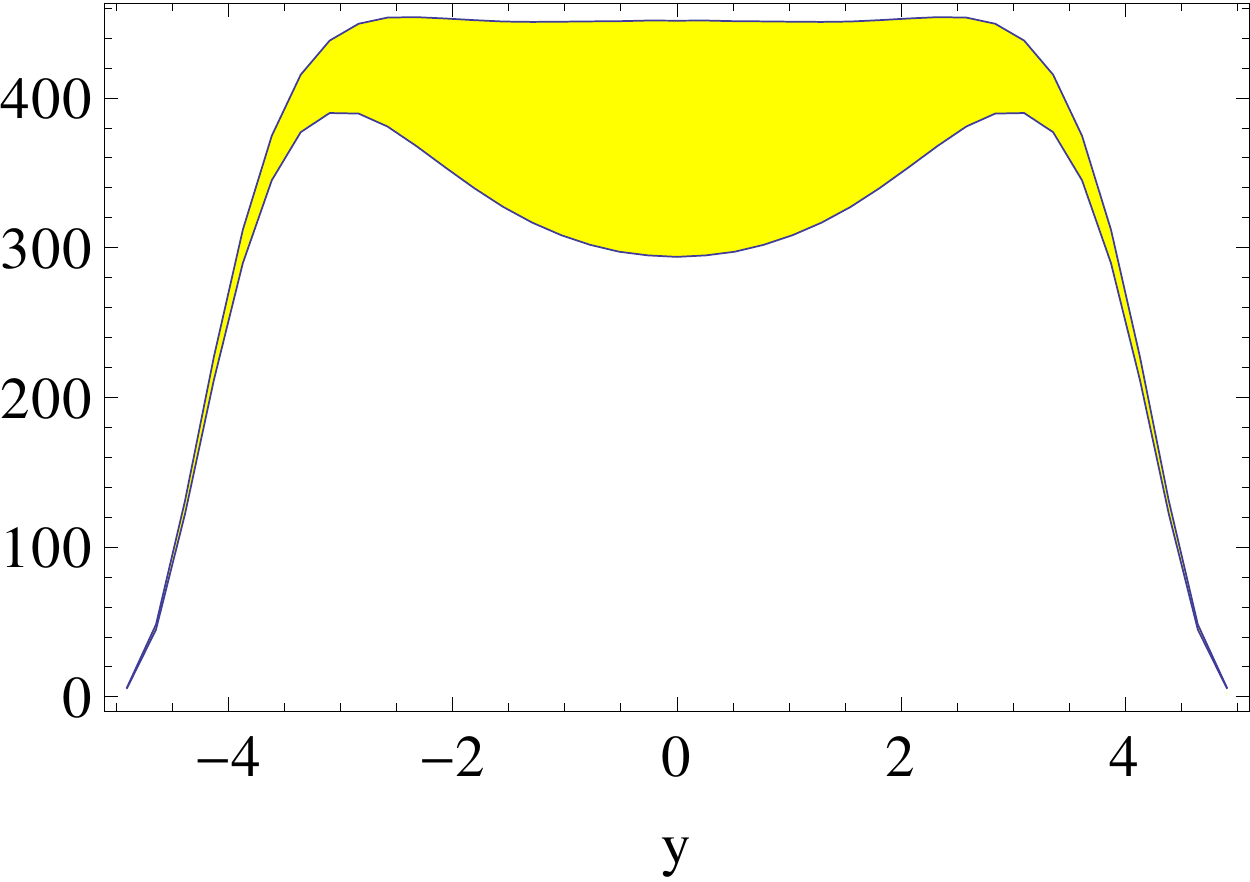}
\caption{a) b) The leading-order (LO) differential cross section ($d\sigma/dy$)
for $W^{+}$production at the Tevatron (2~TeV) and the LHC (14~TeV)
as a function of rapidity. The partonic contributions are also displayed
for $\{u\bar{d},c\bar{s},u\bar{s},c\bar{d}\}$. The vertical scales
are logarithmic.
c) The strange quark contribution (yellow) as a fraction of
the total $d^{2}\sigma/dM/dy$ in pb/GeV for $pp$ to $W^{+}$ production at the LHC for 14~TeV
with CTEQ6.6 using the VRAP program at NNLO. C.f. Ref.~\cite{Kusina:2012vh}
for details.
}
\label{fig:w}       
\end{figure}

To illustrate  the challenges of fitting the nPDFs, in Fig.\ref{fig:rf} we display 
the effective nuclear correction factors $R$ on iron for 
neutral-current charged-lepton ($\ell^\pm$)  DIS and 
charged-current neutrino ($\nu,\bar\nu$) DIS. 
The $R$ values differ, and this uncertainty then feeds into the 
individual flavors such as the strange quark $s(x)$. 
Prior to the LHC data, our knowledge of $s(x)$ came primarily from 
$\nu$Fe DIS, so the uncertainty of the nuclear correction was an important consideration. 
Consequently, improvements in the nuclear correction factor $R$ will 
yield improved proton PDFs. 

At the higher energy scales of the LHC, the (relatively) heavier 
parton flavors (such as the strange) play a more prominent role. 
This is illustrated in Fig.\ref{fig:w} where we find the $c\bar{s}\to W^+$ contribution 
is much larger than at the Tevatron (for example) where the $u\bar{d}$ process dominated. 
We can then turn this issue around and use the LHC data to impose constraints on the strange 
quark.\cite{Aad:2012sb} 

In this instance, the combination of fixed-target $\nu N$ DIS data together with LHC $W/Z$ production data
can improve constraints of both the strange PDF and the nuclear correction factors.

\subsection{Charm and Bottom Quark}

\begin{figure}[ht]
\centering
\includegraphics[width=0.35\textwidth,clip]{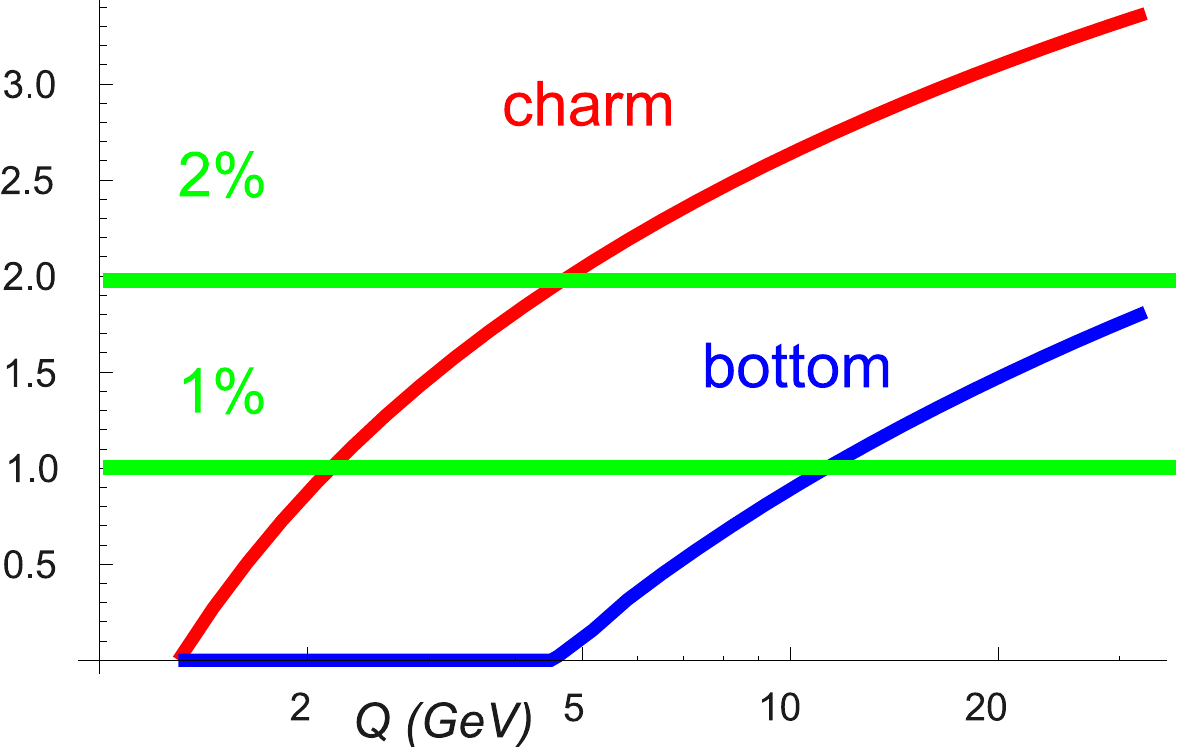}
\hfil
\includegraphics[width=0.35\textwidth,clip]{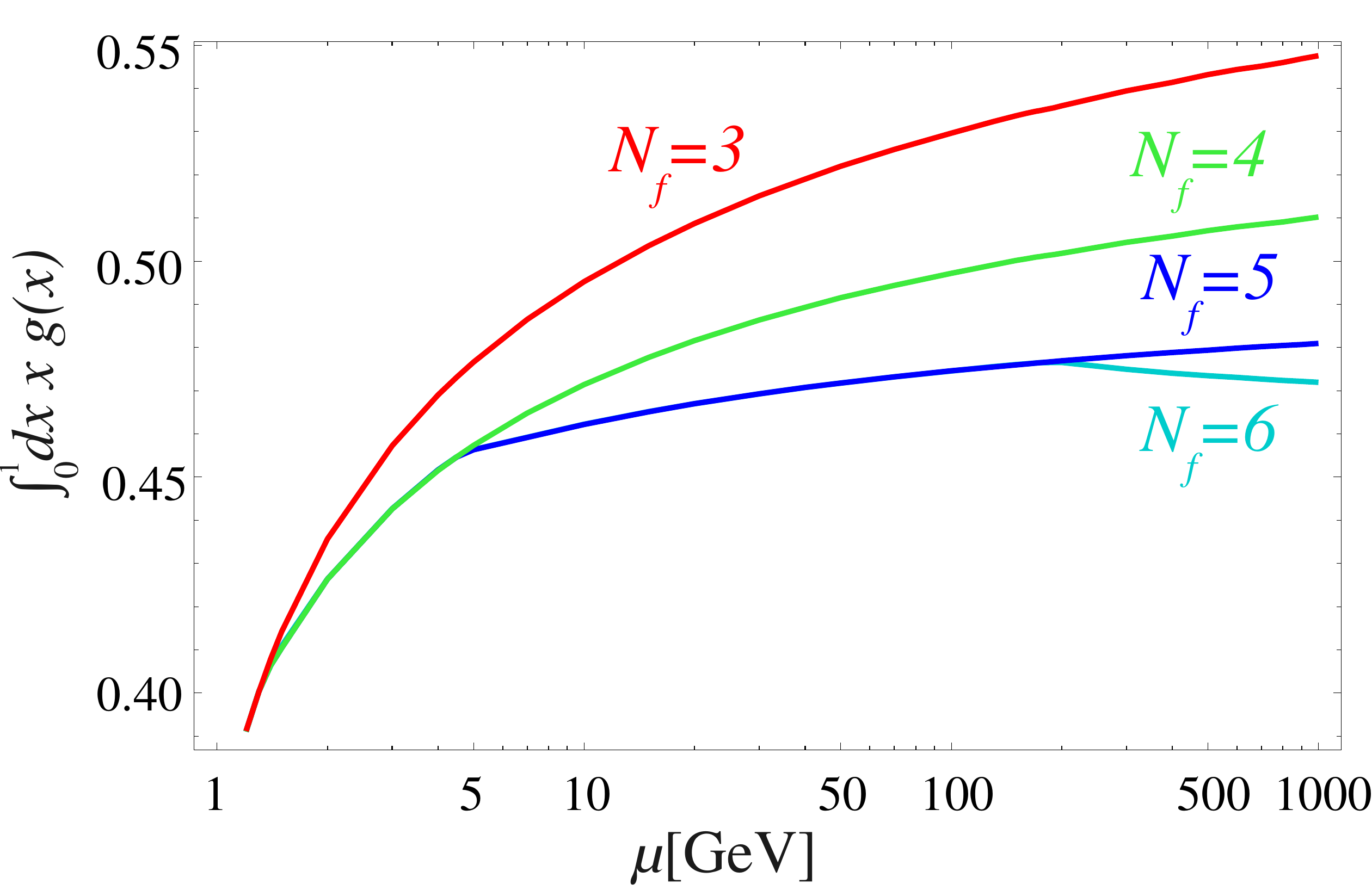}
\caption{
a) Momentum fraction of the charm and bottom PDFs 
as a function of the scale.  
b) Momentum fraction of the gluon for $N_F=\{3,4,5,6 \}$
as a function of the scale. 
}
\label{fig:cb}       
\end{figure}

\begin{figure}[ht]
\centering
\includegraphics[width=0.35\textwidth,clip]{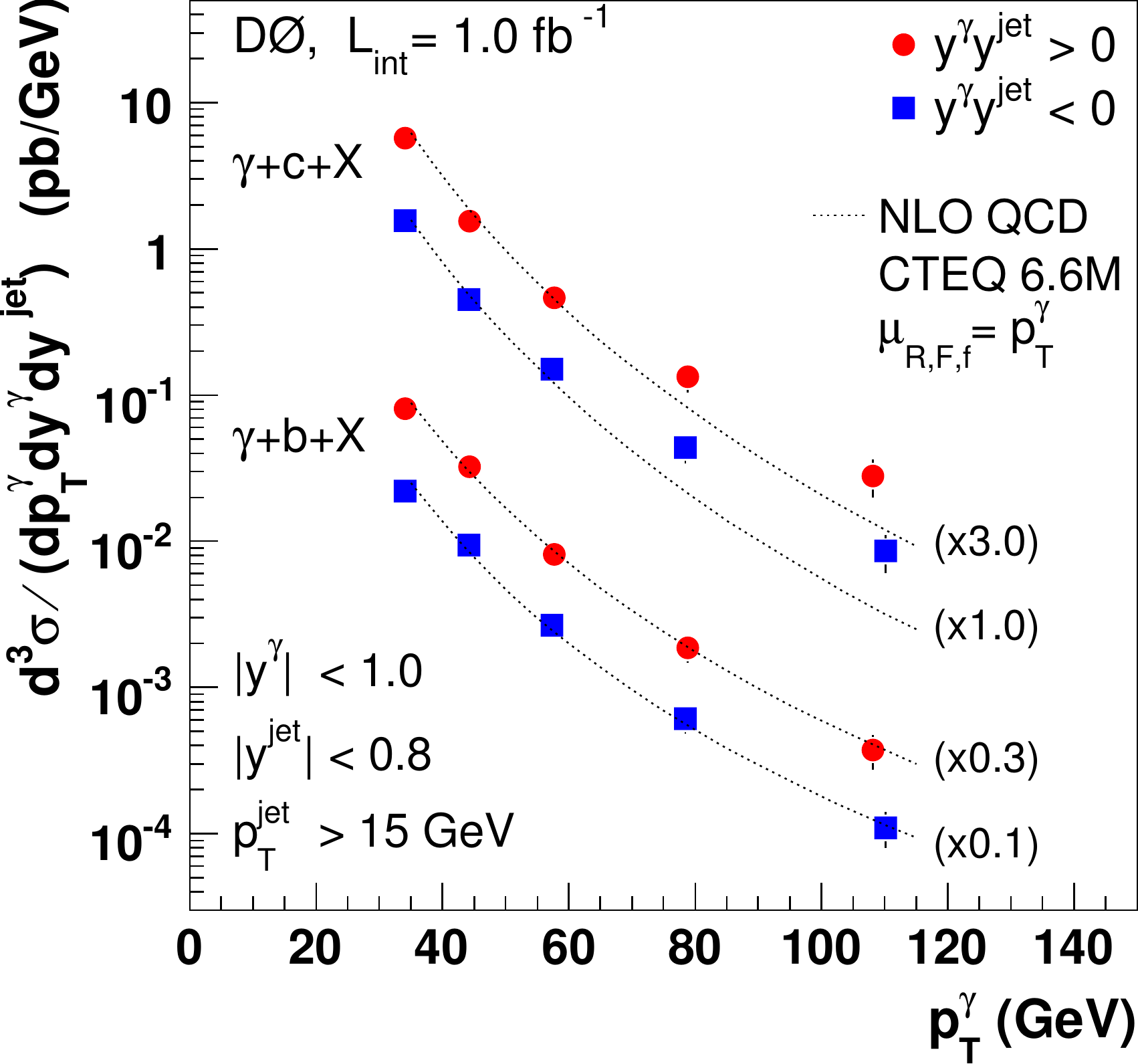}
\hfil
\includegraphics[width=0.40\textwidth,clip]{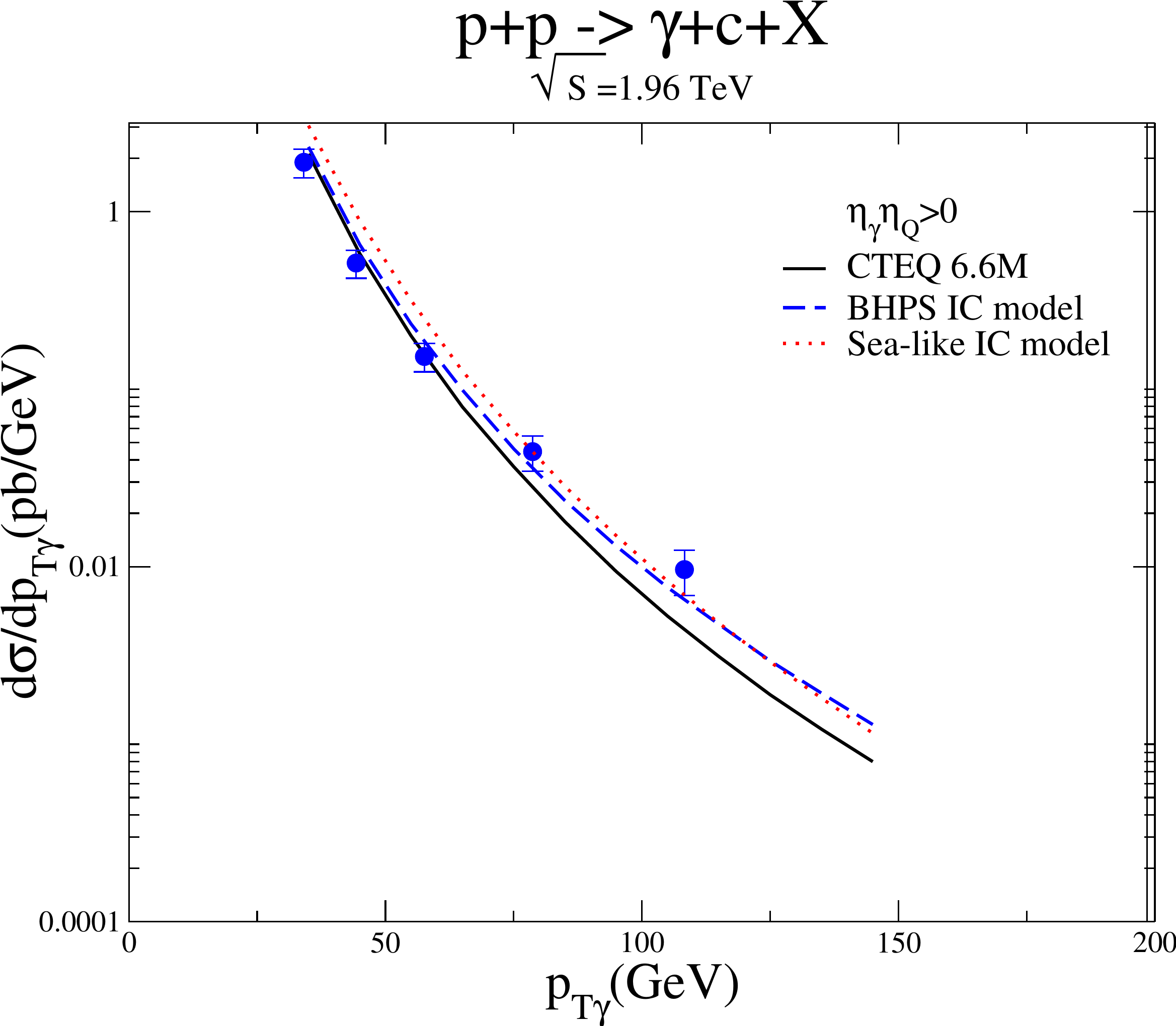}
\caption{
a) Cross section for associated production of $\gamma$ and $c/b$.\cite{Duggan:2009wk}
b) Same with an intrinsic charm component.\cite{Stavreva:2009vi}
}
\label{fig:dirp}       
\end{figure}

Considering the heavier charm and bottom quarks, these enter the PDF 
for energy scales above the quark mass values; as the $c$ and $b$ PDFs 
are primarily produced by gluon 
splitting ($g\to c\bar{c}$ and  $g\to b\bar{b}$), these affect the momentum 
fraction carried by the gluon ({\it c.f.}, Fig.\ref{fig:cb}).

While we generally assume that there is no primordial component of the $c$ and $b$ PDFs
for energy scales below the quark mass, it has been postulated that there could be 
an additional ``intrinsic'' component which would be added to the usual ``extrinsic'' 
component (which comes from gluon splitting). There are a number of measurements 
where the ``intrinsic'' component could manifest itself; see Ref.~\cite{Brodsky:2015fna} for a recent review.

Fig.\ref{fig:dirp}-a shows the data for $\gamma+c\to X$ is above the theory prediction at 
the largest $p_T^{\gamma}$ values; an additional ``intrinsic'' component 
(Fig.\ref{fig:dirp}-b) can shift the theory in the direction of the data at large  $p_T^{\gamma}$. 
While this observation is intriguing, additional precision and analysis is required
to definitively claim evidence of an ``intrinsic'' component.

On a related topic, 
recent theoretical work demonstrated that (to a good approximation)
the evolution of the intrinsic heavy quark component is governed by
non-singlet evolution equations which can be decoupled from the
evolution of the remaining PDFs.\cite{Lyonnet:2015dca} The observation
allows an (approximate) intrinsic component to easily be included with
any calculation without the need for a specially evolved PDF set; this
greatly simplifies our ability to search for, and place constraints
upon, intrinsic charm and bottom components of the nucleon.

\subsection{Higher Order Calculations}

\begin{figure}[ht]
\centering
\includegraphics[width=0.35\textwidth,clip]{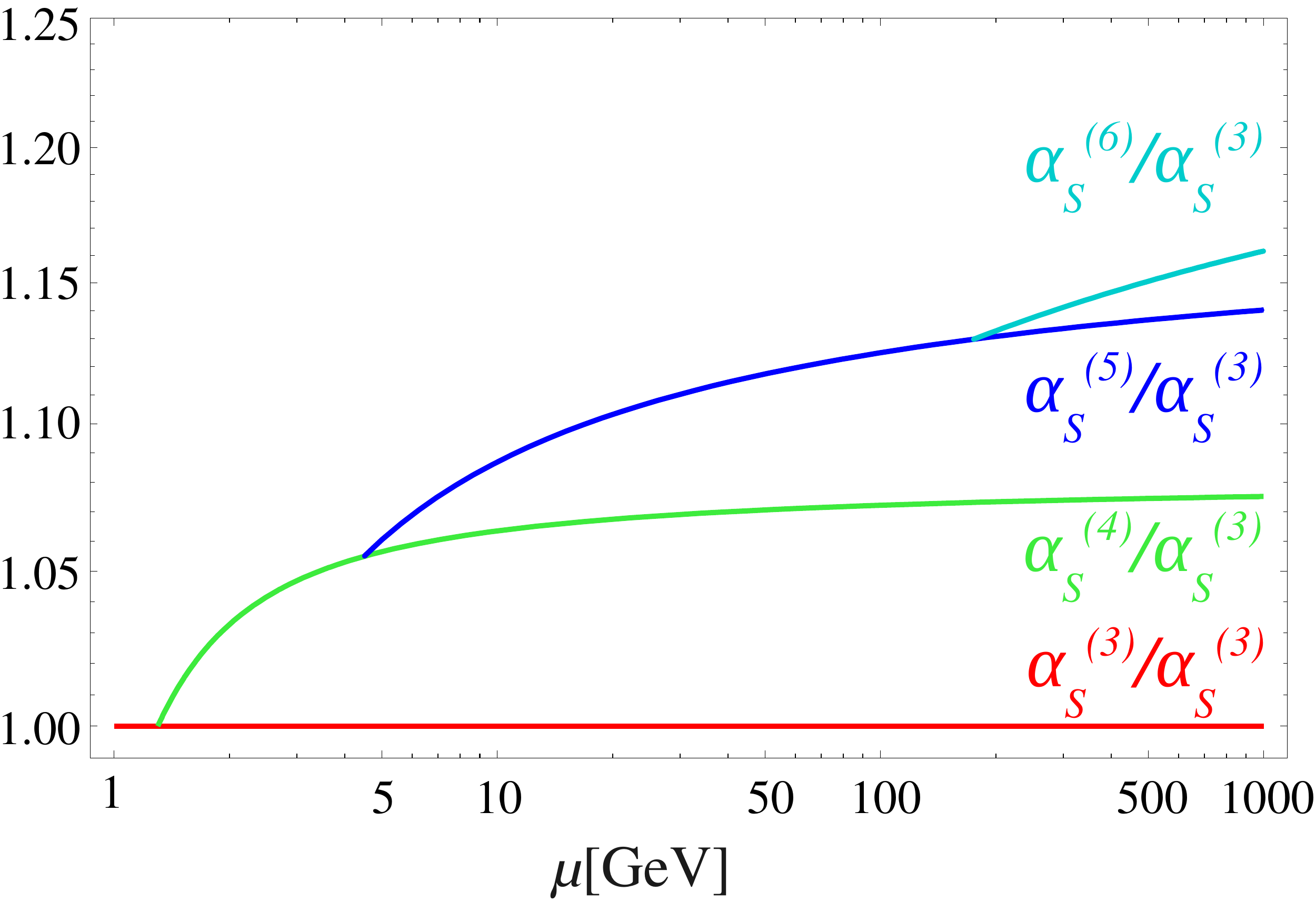}
\hfil
\includegraphics[width=0.35\textwidth,clip]{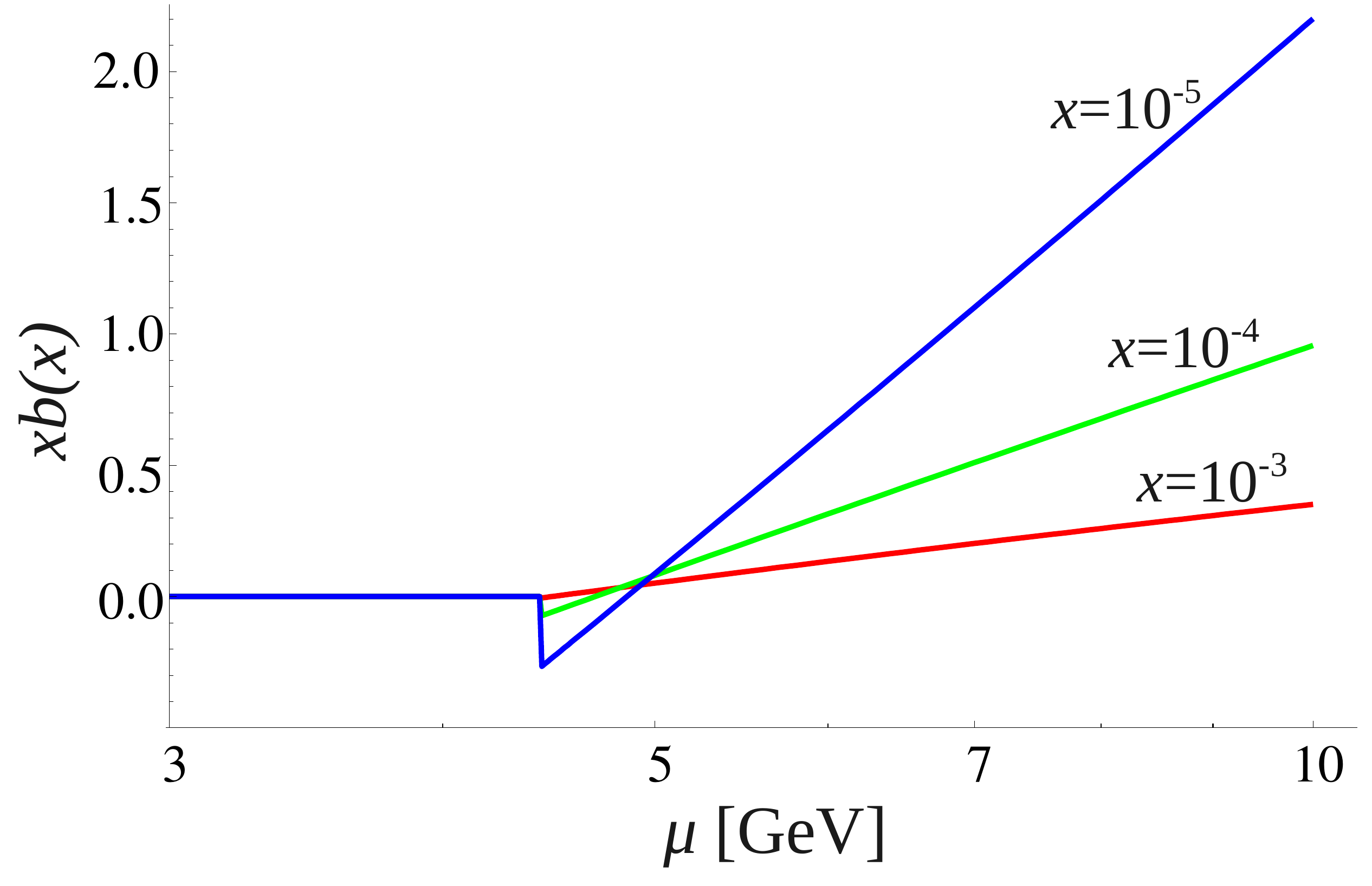}
\caption{
a) Ratios of $\alpha_s(\mu)$ for $N_F=\{3,4,5,6\}$.  
b) The discontinuity in the $b$-quark PDF $f_b(x,\mu)$ at NNLO {\it vs.} $\mu$ for
$x=\{10^{-3},10^{-4},10^{-5}\}$.
}
\label{fig:hiord}       
\end{figure}

Recent advances in the automation of higher order calculations have tremendously
increased the breadth of processes which are available at NNLO and beyond. 
As we move to these higher orders,  our PDFs and $\alpha_s(\mu)$ 
functions are discontinuous\footnote{$\alpha_s(\mu)$ has discontinuities at order 
$\alpha_s(\mu)^3$, but these are not apparent in Fig.\ref{fig:hiord}-a.}
 when we shift the number of flavors from $N_F$ to  $N_F+1$; 
 {\it c.f.} Fig.\ref{fig:hiord}.
An elegant property of QCD is that despite the (fixed order) discontinuity of the PDFs and  $\alpha_s$,
the discontinuity of any physical quantity  will systematically decrease order by order
in perturbation theory; that is to say, at order $k$ we have the relation: 
$\sigma \simeq f_{N_F} \otimes  \hat\sigma_{N_F} = f_{N_F+1} \otimes  \hat\sigma_{N_F+1} +{\cal O}(\alpha_s^{k+1}).$ 
This is certainly a remarkable result.

\subsection{Incisive Analysis Tools}

\begin{figure}[ht]
\centering
\includegraphics[width=0.55\textwidth,clip]{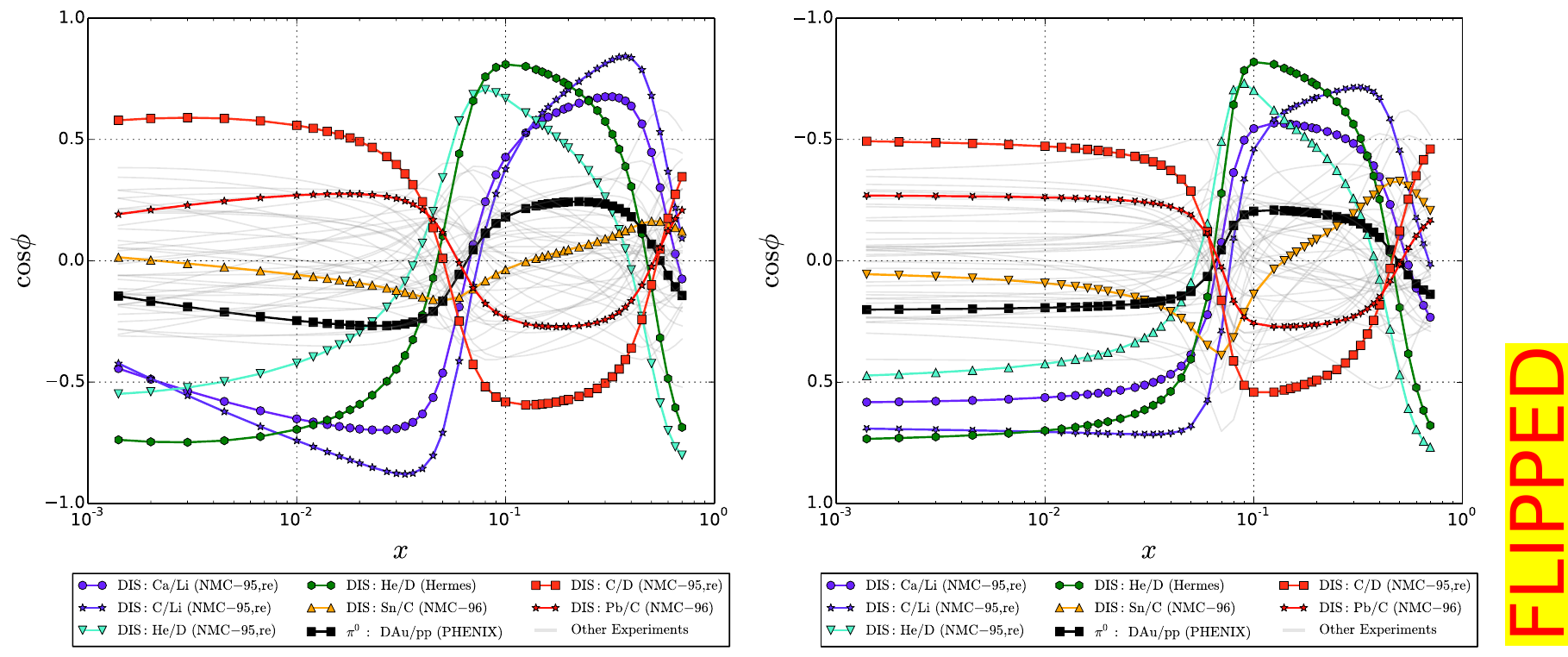}
\caption{
Correlation measures ($\cos\phi$) for $u$-valence  (left) and $d$-valence (right) for lead at $Q = 10\GeV$.
Eight selected experiments are highlighted with symbols. To emphasize the anti-correlation
between $u$-valence  and $d$-valence  we have flipped the $d$-valence  plot vertically. 
See Ref.\cite{Kovarik:2015cma} for the original figures and additional details.
}
\label{fig:corr}       
\end{figure}

Finally, as we advance our analysis methods it is important to develop new
tools and metrics that can insightfully  and effectively convey the essence of the
physics.
In Fig.\ref{fig:corr} we display the correlation cosine $\cos\phi$
{\it vs.} $x$ for a variety of experiments. The plot effectively conveys
which experiments influence the PDFs in different $x$ regions, and how
they are correlated. To emphasize this point we compare the $u$-valence
(left) with the flipped $d$-valence figure (right); by flipping the
figure we see that $u$-valence and $d$-valence are anti-correlated.
Thus, as the fit works to find a minimum $\chi^2$ value, it has to
balance these experimental data sets which are pulling the  $u$-valence and
$d$-valence in opposite directions.

\subsection{Conclusions}

Recent theoretical developments provide improved 
precision for both proton and nuclear targets across a wide kinematic range. 
Further improvements will rely on a combination of additional theoretical advances, 
and a broad range of experimental measurements. The prospect of future data from EIC and LHeC facilities 
would be a valuable complement to the LHC and advance both our search for new physics and our understanding of QCD.



\end{document}